\documentclass[aps,amsmath,twocolumn,amssymb,titlepage,superscriptaddress,10pt]{revtex4-1}
\usepackage[T1]{fontenc}
\usepackage[utf8]{inputenc}
\usepackage{amsmath}
\usepackage{braket}
\usepackage{amsfonts}
\usepackage{comment}
\usepackage{graphicx}
\usepackage{hyperref}
\usepackage[capitalize]{cleveref}
\usepackage{amssymb}
\usepackage{amsmath}
\usepackage{mathtools}
\usepackage{todonotes}

\DeclareMathOperator{\Tr}{Tr}

\newcommand{\Ham}{H}
\newcommand{\bL}{\beta}
\newcommand{\sumtwo}[2]{\sum_{\substack{#1 \\ #2}}}
\newcommand{\bbC}{{\mathbb C}}
\newcommand{\bbE}{{\mathbb E}}
\newcommand{\bbP}{{\mathbb P}}
\newcommand{\bbS}{{\mathbb S}}
\newcommand{\bbZ}{{\mathbb Z}}
\newcommand{\caB}{{\mathcal B}}
\newcommand{\caL}{{\mathcal L}}

\begin{document}
\title{Poisson-Dirichlet distributions and  weakly first-order \\ spin-nematic phase transitions}
\author{Nils Caci}
\affiliation{Institute for Theoretical Solid State Physics, JARA FIT and JARA CSD, RWTH Aachen University, 52056 Aachen, Germany}
\author{Peter Mühlbacher}
\affiliation{Department of Mathematics, University of Warwick, Coventry CV4 7AL, United Kingdom}
\author{Daniel Ueltschi}
\affiliation{Department of Mathematics, University of Warwick, Coventry CV4 7AL, United Kingdom}
\author{Stefan Wessel}
\affiliation{Institute for Theoretical Solid State Physics, JARA FIT and JARA CSD, RWTH Aachen University, 52056 Aachen, Germany}

\begin{abstract}
 We provide a quantitative  characterization of generic weakly first-order thermal phase transitions out of planar spin-nematic states in three-dimensional spin-one quantum magnets, based on 
 calculations using  Poisson-Dirichlet distributions (PD) within a universal loop model formulation, combined with large-scale quantum Monte Carlo calculations. In contrast to earlier claims, the thermal melting of the nematic state is not  continuous, instead  a weakly first-order transition is identified from  both thermal properties and the distribution of the nematic order parameter. Furthermore, based on PD calculations, we obtain exact results for the  order parameter distribution and  Binder cumulants at the discontinuous melting transition. Our findings establish the thermal melting of planar spin-nematic states as a generic platform for quantitative approaches to  weakly first-order phase transitions in quantum systems with a continuous SU(2) internal symmetry. 
    
\end{abstract}

\maketitle


The theory of phase transitions is fundamental to  modern approaches to many-body systems and quantum matter. In particular continuous phase transitions are a central topic in various areas of physics in view of the universality of critical phenomena. More recently, also
weakly first-order phase transitions, i.e., discontinuous phase transitions with correlations lengths well beyond the lattice  scale, became a topical subject in condensed matter research and beyond~\cite{Kaplan2009,Nahum2015,Wang2017,Gorbenko2018,Gorbenko2018II,Ma2019,Nogueira2019,Iino2019,Demidio2021}. 
Different renormalization group (RG) scenarios explain the emergence of quasi-scaling near weakly first-order phase transitions: In the "tuning" mechanism, the RG-flow passes near a real infrared fixed-point in theory space;
in the case of "walking", the RG-flow passes between two fixed-points at complex couplings, associated to the collapse of two real fixed-points~\cite{Kaplan2009,Gorbenko2018}.
An example, for which the latter scenario  can be demonstrated explicitly,  is the $q$-state Potts model with a discrete $Z_q$ symmetry, featuring  weakly first-order transitions for $q>4$ in two dimensions~\cite{Wu1982,Nauenberg1980,Cardy1980,Gorbenko2018II}.
Similar ideas  relate to the hierarchy generation in four-dimensional gauge theories of high-energy physics within the framework of walking as a slowly running coupling constant at intermediate energy scales~\cite{Holdom1981,Yamawaki1986,Appelquist1986,Gorbenko2018}.

For quantum  many-body systems weakly first-order transitions are also central to some recently proposed interpretations of the deconfined quantum criticality (DQC) scenario \cite{Senthil2004,Sandvik2007,Senthil2004b,Shao2016,Ma2018} in terms of walking, fixed-point annihilation,  and complex fixed points~\cite{Nahum2015,Wang2017,Ma2020,Nahum2020}: For DQC, quantum many-body systems are considered with continuous internal symmetries, such as U(1) or spin SU(2), for which the anticipated DQC points separate ordered regions with non-compatible symmetry-breaking patterns. The debate is still ongoing, regarding the true nature of the quantum phase transitions observed in various DQC designer models, as well as their relation to specific quantum materials~\cite{Yang2022,Cui2022}. In view of these developments, it is crucial to establish quantum  systems in which  weakly first-order transitions can be robustly demonstrated and  exact results for the properties at the transition point can be provided by insightful approaches.

Here, we uncover weakly first-order transitions in spin-one quantum magnets on the three-dimensional cubic lattice with SU(2) symmetric interactions. By large-scale quantum Monte Carlo (QMC) simulations, we establish that in contrast to earlier claims of a continuous  transition~\cite{Harada2002}, the planar spin-nematic (ferroquadrupolar) phase that emerges  in this system melts across a weakly first-order transition. Its discontinuous nature becomes apparent (in both the thermodynamic properties as well as the order parameter distribution) only on sufficiently large length scales, beyond those accessed previously~\cite{Harada2002}. In contrast to the case of the  Potts model and designer models of DQC, in this system the weakly first-order transition takes place between a paramagnet  and a low-temperature ordered phase that breaks a continuous internal SU(2) symmetry.
 Moreover, we show how calculations based on  PD  within a universal loop-model formulation of the spin-one lattice model can be used to derive the exact order parameter distribution in the spin-nematic phase as well as order parameter Binder cumulants at the transition point, thereby providing us with a quantitative characterization of this weakly first-order thermal order-disorder transition. We note that the spin-one material NiGa${}_2$S${}_4$ is 
 a candidate system for the 
 planar spin-nematic state considered here~\cite{Nakatsuji2005,Nambu2006,Tsunetsugu2006,Laeuchli2006,Bhattacharjee2006,Valenntine2020}.

\textit{Model and planar spin-nematic.}
To stabilize the planar spin nematic state,
we consider the generic SU(2)-symmetric 
spin-one Hamiltonian, i.e., with both bilinear and biquadratic  interactions, on a simple
cubic lattice $\Lambda$,
\begin{equation}
\label{def ham}
    \Ham = - J \sum_{\langle i,j\rangle \in\caB_\Lambda}\left[  \: u \,( \mathbf{S}_i \cdot \mathbf{S}_j )+   v \, (\mathbf{S}_i \cdot \mathbf{S}_j)^2\: \right],
\end{equation}
with $|\Lambda|=L^3$ sites, and a sum  over the (nearst-neighbor) bonds $\caB_\Lambda$ of $\Lambda$ (with periodic boundary conditions).  It is convenient to fix $v=1$ and keep $u$ as a free parameter.  Alternatively,  an angular parametrization, $u=\cos(\phi)$,  $v=\sin(\phi)$ can be used. In any case, we fix $J=1$.
For $u\in(0,1)$ [i.e., $\phi\in(\pi/4,\pi/2)$], $\Ham$
harbors an extended planar spin-nematic phase~\cite{Tanaka2001,Harada2002,Batista2004,Toth2012,Fridman2013,Ueltschi2015}, in which  magnetic fluctuations are constrained to the plane perpendicular to a director $\vec a \in P\bbS^2$, the projective sphere, i.e., $\vec a$ is identified with $-\vec a$. Each director
corresponds to an extremal Gibbs state $\langle \cdot \rangle_{\vec a}$~\cite{Ueltschi2015}.
The symmetric Gibbs state at inverse temperature $\beta=1/T$, $\langle \cdot \rangle_{\bL} = \Tr \cdot \:e^{-\beta \Ham}/Z$, $Z=\Tr e^{-\beta \Ham}$,
in the infinite-volume limit,  then has the decomposition~\cite{Ueltschi2015}
    $\lim_{L\to\infty} \langle \cdot \rangle_{\bL} = \int_{P\bbS^2} \langle \cdot \rangle_{\vec a} \, d\vec a.$
Here, $d\vec a$ denotes the uniform probability measure on $P\bbS^2$. In general $\langle \cdot \rangle_{\vec a}$ depends on $\beta$ (for small $\beta$ the Gibbs state is unique and $\langle \cdot \rangle_{\vec a}$ does not depend on $\vec a$). A suitable local operator to detect nematic order is
    $Q_i = (S_i^{\rm z})^2 - \tfrac23$,
and we denote by $n^*$  the "spontaneous nematisation" in the $z$ direction,
    $n^* = \langle Q_i \rangle_{\vec e_{\rm z}}$,
where $i$ is any site. From the PD formulation introduced below, it  follows that  $n^*<0$~\cite{SM}. In contrast to the axial nematic state that appears, e.g.,  for $\Ham$ with classical spins at $u=0$~\cite{Angelescu1982,Biskup2003},  the planar nematic phase is characterized by the {\it minimization}  of the fluctuations in the plane perpendicular to the director, and 
$\langle \cdot \rangle_{\vec a} = \lim_{h \to 0+} \lim_{L\to\infty} \langle \cdot \rangle_{\Ham + h \sum_{i \in \Lambda} (\vec a \cdot \vec S_i)^2}$ (notice the "+" sign in front of $h$)~\cite{Ueltschi2015}. This is a genuine quantum mechanical phenomenon, related to the $m=0$ state of the spin-one variables in this system. 
For $u=0$ and $u=1$, the model exhibits 
an enhanced SU(3) symmetry and ferromagnetic low-temperature order~\cite{Ueltschi2013,Ueltschi2015}. In the following, we study the properties of the model $\Ham$ at finite  $T$, in particular the nature of the  thermal melting of the spin-nematic state and its quantitative description. 

\begin{figure}[t]
    \centering
    \includegraphics[width=84mm]{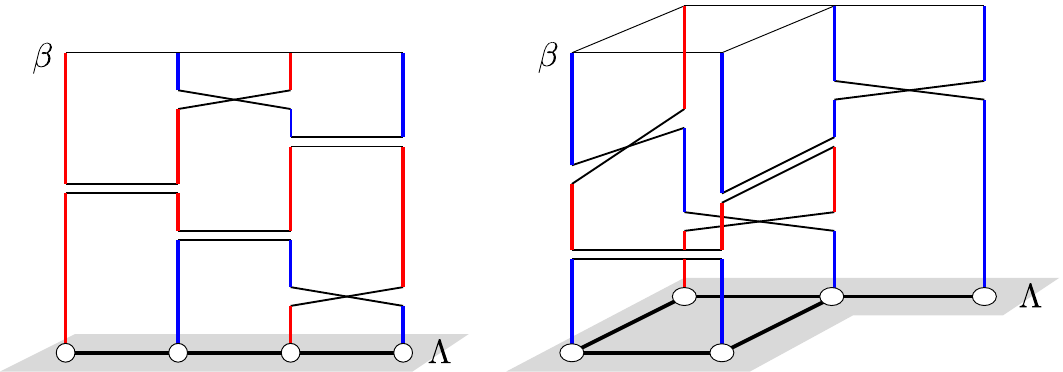}
    \caption{Illustration of the loop model  on different small lattices (for $\Ham$, $\Lambda$ is the cubic lattice).}
    \label{fig loops}
\end{figure}

\textit{Loop model and PD predictions.}
Loop models involve one-dimensional objects "living" in $d$-dimensional space. Phases may occur where loops of diverging lengths are present. It was recently observed in \cite{Goldschmidt2011} that the joint distribution of the lengths of long loops displays universal behaviour: It is always given by the stationary distribution of a split-merge process, which is PD characterized by a real number, the PD parameter $\theta$ (cf.~\cite{SM} for a basic introduction to PD and split-merge processes). We denote the corresponding distribution by PD($\theta$).
It is possible to derive a loop model representation for $\Ham$ 
using the Trotter or Duhamel formul\ae{}  
for the Gibbs operator ${\rm e}^{-\beta \Ham}$. It is restricted to $u\in [0, 1]$ (outside this domain, the representation involves negative weights). This combines representations due to T\'oth \cite{Toth1993} and to Aizenman and Nachtergaele \cite{Aizenman1994} and was proposed in \cite{Ueltschi2013}. The latter article contains a detailed derivation.
The resulting  representation is illustrated in Fig.\ \ref{fig loops}. On top of each bond of the spatial lattice $\Lambda$ is the "time" interval $[0,\beta]$. In each interval is an independent Poisson point process where "crosses" occur with intensity $u$ and "double bars" occur with intensity $1-u$. One then defines the loops as the closed trajectories obtained by moving vertically, and jumping on the neighbouring site when encountering a cross or a double bar. If it is a cross, one continues in the same vertical direction, while if it is a double bar, one changes the vertical direction. The role of the loops is  twofold:(i) they affects the probability of the loops because of a factor $3^{\# {\rm loops}}$, (ii)  quantum correlations are given by loop correlations. The relation between quantum spins and loops concerns the partition function via
\begin{equation}
\begin{split}
    Z&= e^{2\beta |\caB_\Lambda|} \sum_{k,\ell=0}^\infty \frac{(1-u)^k u^\ell}{k! \, \ell!} \\
    &\times \sumtwo{b_1,\dots,b_k}{c_1,\dots,c_\ell} \int_0^\beta ds_1 \dots ds_k dt_1 \dots dt_\ell \, 3^{|\caL(\omega)|}.
\end{split}
\end{equation}
Here, $\omega$ denotes a configuration in terms of  $b_1,\dots,b_k \in \caB_\Lambda$  ($c_1, \dots, c_\ell \in \caB_\Lambda$), the bonds  corresponding to double bars (crosses), and $s_1, \dots, s_k \in [0,\beta]$ ($t_1,\dots,t_\ell \in [0,\beta]$), the times at which double bars (crosses) occur. $\caL(\omega)$ denotes the set of loops.
%

Furthermore, we obtain for
the characteristic function of the "nematic histogram", i.e., 
the distribution function $\rho_Q$ of the ferroquadrupolar operator $Q=\frac1{|\Lambda|} \sum_{i \in \Lambda} Q_i$ in the Gibbs state $\langle \cdot \rangle_{\bL}$, 
the identity (for any $k \in \bbC$) 
\begin{equation}
\label{spins to loops}
    \Bigl\langle e^{ikQ} \Bigr\rangle_{\bL} \!\!
    = \biggl\langle \prod_{\gamma \in \caL(\omega)}\!\! \bigl( \tfrac13 e^{-\frac23 \frac{ik}{|\Lambda|} \ell(\gamma)} + \tfrac23 e^{\frac13 \frac{ik}{|\Lambda|} \ell(\gamma)} \bigr) \!\!\biggr\rangle^\text{\!\!loops}_{\bL},
\end{equation}
where the length $\ell(\gamma)$ of the loop $\gamma$ is defined as the number of sites traversed by the loop at time 0, and  $\langle\cdot\rangle^\text{loops}_{\bL}$ denotes the expectation with respect to the loop measure above.
This measure can be viewed as the invariant measure of a Markov process, involving the insertion  and removal of double bars and crosses~\cite{Ueltschi2015,Ueltschi2017}, as detailed in ~\cite{SM} (note that this process would be too slow to use in simulations.)
Any new cross or double bar between two loops causes them to merge. When $u \in (0,1)$, a subtle phenomenon occurs: a new cross or double bar may either cause a loop to split, or reorganise it without splitting it (this is akin to $0 \leftrightarrow 8$); either occurs with probability $\frac12$. 
The lengths of macroscopic loops can be shown to satisfy an effective split-merge process, and the invariant distribution is PD(3/2) \cite{Tsilevich2000,Diaconis2004,Goldschmidt2011}.
For $u = 0$ or $u = 1$, the subtle phenomenon above does not occur; splits then happen at twice the rate, and $\theta = 3$.

The PD conjecture~\cite{Goldschmidt2011,Ueltschi2015} states that, as $L\rightarrow \infty$, we can replace the expectation in the loop model by the expectation with respect to PD($\theta$), scaled by a number $\eta = \eta(u,\beta)\in [0,1]$ that represents the fraction of sites in long loops at imaginary time 0. This can be used to calculate the characteristic function of $\rho_Q$ explicitly~\cite{SM}:
\begin{equation}
\label{expression for exp(ikQ)}
    \lim_{L\to\infty} \Bigl\langle e^{{ik} Q} \Bigr\rangle_{\bL} = e^{-\frac23 ik \eta} \sum_{r=0}^\infty \frac{\Gamma(\frac32)}{\Gamma(r+ \frac32)} (ik \eta)^r.
\end{equation}
Inverse Fourier transform finally gives~\cite{SM}
\begin{equation}
     \rho_Q(s) = \begin{cases} \frac1{2\sqrt\eta \sqrt{\frac13 \eta - s}} & \text{if } -\frac23 \eta \leq s \leq \frac13 \eta, \\ 0 & \text{otherwise.} \end{cases}
\end{equation}

\begin{table}[t]
\centering
 \begin{tabular}{c|cccc}
 & \quad  $\theta$ \quad  &  \quad $\langle Q^2 \rangle_{\bL}$ \quad & $\quad  \langle Q^3 \rangle_{\bL}$ \quad & $\quad \langle Q^4  \rangle_{\bL}$ \quad   \\ 
 \hline
 $u \in (0, 1)$ & 3/2 & $\frac4{45} \eta^2$ & $-\frac{16}{27\cdot 35} \eta^3$ & $\frac{16}{27\cdot 35} \eta^4$  \\
 $u \in \{ 0, 1\}$ & 3 & $\frac1{18} \eta^2$ & $-\frac1{135} \eta^3$ & $\frac1{135} \eta^4$  \\
 \end{tabular}
 \caption{PD results for the moments of $Q$.}\label{Table:Qn}
\end{table}
 
From here, we can calculate the moments  
$\langle Q^n \rangle_{\bL}$ in the nematic phase. It is more insightful however, to derive them from the loop representation directly, using the PD conjecture to write them all in terms of the single unknown variable $\eta$.
Cumulant ratios, such as $U_Q=1-\frac13\langle Q^4 \rangle_{\bL}/\langle Q^2 \rangle_{\bL}^2$, are then given by ratios that do not depend on $\eta$ any more. 
We provide the calculations in some details since they cannot be found in the literature, to the best of our knowledge~\cite{SM}. The identities for the 2nd, 3rd and 4th moment that are exact in the infinite-volume limit read
\begin{equation}
    \begin{split}
        &\langle Q^2 \rangle_{\bL} = \tfrac29 \bbP_{\bL}^{\rm loops} [ i_1, i_2 \text{ in same loop}], \\
        &\langle Q^3 \rangle_{\bL} = -\tfrac2{27} \bbP_{\bL}^{\rm loops} [ i_1, i_2, i_3 \text{ in same loop}],\\
        &\langle Q^4 \rangle_{\bL} = \tfrac2{27} \bbP_{\bL}^{\rm loops} [ i_1, i_2, i_3, i_4 \text{ in same loop}] \\
     &+ \tfrac4{27} \bbP_{\bL}^{\rm loops} [ i_1, i_2 \text{ in same loop}, i_3, i_4 \text{ in other loop}].
    \end{split}
\end{equation}
Here $i_1, i_2, i_3$ and $i_4$  are sites that are very distant from one another. 
Since the sites are distant, it is necessary that they belong to long loops in order to have a chance to be in the same loop. We can then use the PD conjecture to obtain the probability $\bbP_{\bL}^{\rm loops} [ i_1, \dots, i_n \text{ in same loop}]$, that $i_1,...,i_n$ belong to the same loop,
in term of the probability that, if we choose a random partition of $[0,1]$ according to PD($\theta$), and $n$ independent points in $[0,1]$, all $n$ points are in the same partition element~\cite{SM}:
\begin{equation}
\begin{split}
        &\bbP_{\bL}^{\rm loops} [ i_1, \dots, i_n  \text{ in same loop}]  \nonumber\\
        &=\eta^n \bbP_{{\rm PD}(\theta)} [n \text{ random points in same partition element}]\nonumber\\
        &= \eta^n \frac{\Gamma(1+\theta) \Gamma(n)}{\Gamma(n+\theta)},
\end{split}
\end{equation}
and similarly
\begin{equation}
\begin{split}
&\bbP_{\bL}^{\rm loops} [i_1, i_2 \text{ in same loop}, i_3, i_4 \text{ in other loop}] \nonumber \\
&= 2\eta^4 \sum_{k<\ell} \bbP_{{\rm PD}(\theta)} [i_1, i_2 \text{ in $k$th element}, i_3, i_4 \text{ in $\ell$th el.}] \nonumber \\
&= 2\eta^4 \frac{\theta \Gamma(1+\theta)}{\Gamma(4+\theta)}.
\end{split}
\end{equation}
The resulting moments are given in Tab.~\ref{Table:Qn}. We obtain an $\eta$-independent  value  $U^-_Q=2/7$ for the Binder cumulant in the thermodynamic limit  within the planar spin-nematic phase, and $\eta$-independent values for the ratios of the moments towards the SU(3) end points, such as
$\lim_{u\to 0^+}\langle Q^2 \rangle_{\bL}(u)/\langle Q^2 \rangle_{\bL}(u=0)=
\lim_{u\to 1^-}\langle Q^2 \rangle_{\bL}(u)/\langle Q^2 \rangle_{\bL}(u=1) =8/5$.
The moments of $Q$ can also  be  calculated using  symmetry breaking extremal states~\cite{SM}, for which however the heuristics is more subtle and the result may be uncertain.

\begin{figure}[t]
    \centering
    \includegraphics{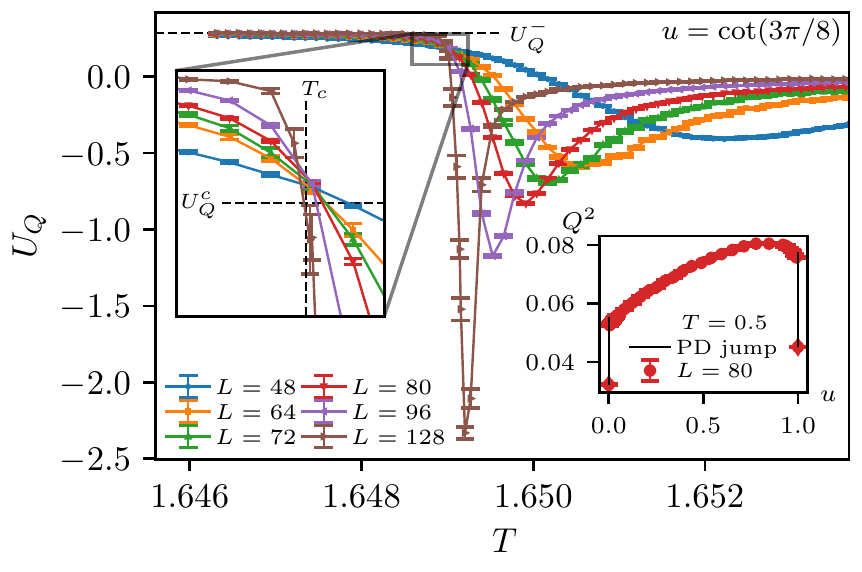}
    \caption{Temperature dependence of the Binder cumulant $U_Q$ near the phase transition for different system sizes at $u=\cot(3\pi/8)$ from QMC. The PD-based predictions $U_Q^-$ in the ordered phase and $U_Q^c$ at the transition temperature $T_c$ are indicated by dashed lines. The right inset shows the $u$-dependence of the second moment $\langle Q^2 \rangle_{\bL}$.}
    \label{fig:U}
\end{figure}

\textit{Comparison to QMC.}
\begin{figure}[t]
    \centering
    \includegraphics{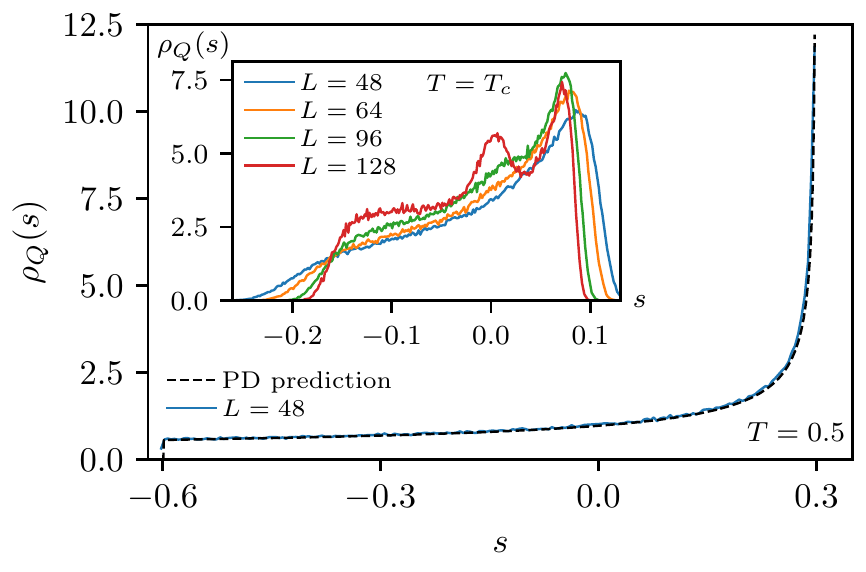}
    \caption{Comparison of the nematic histogram obtained from QMC at $u=\cot(3\pi/8)$ with the PD prediction for $\eta=0.8981(9)$. The inset show the histogram for various $L$ at the transition temperature $T_c=1.64900(1)$.}
    \label{fig:PQ}
\end{figure}
We verify the above results, obtained from the PD conjecture, by making use of unbiased large-scale QMC simulations, based on the stochastic series expansion~\cite{Sandvik1999,Syljuasen2002}. Figure~\ref{fig:PQ} compares the PD prediction for $\rho_Q$ with the nematic histogram obtained using QMC simulations for  $u=\cot(\phi=3\pi/8)=0.41412...$, i.e., at the center of the  spin-nematic regime in  the  angular parametrization of $\Ham$, at a low temperature of $T=0.5$ in the ordered phase (similar results are obtained for other values of $u$). We observe a remarkable agreement between the nematic histogram and the PD prediction.
In Fig.~\ref{fig:U}, we show the thermal evolution of $U_Q$ and find that within the ordered phase, the QMC data converges towards the PD prediction upon increasing the system size ($U_Q$ converges to 0 in the paramagnetic regime). We also examine in the inset of Fig.~\ref{fig:U} the $u$-evolution of 
$\langle Q^2 \rangle_{\bL}$ at a fixed low temperature $T=0.5$. 
We observe an explicit $u$-dependence of the 2nd moment (reflected by the $u$-dependence of $\eta$ in the PD prediction), as well as the agreement 
in the relative size of its jump to both SU(3) end points with the PD prediction.  

\begin{figure}[t]
    \centering
    \includegraphics{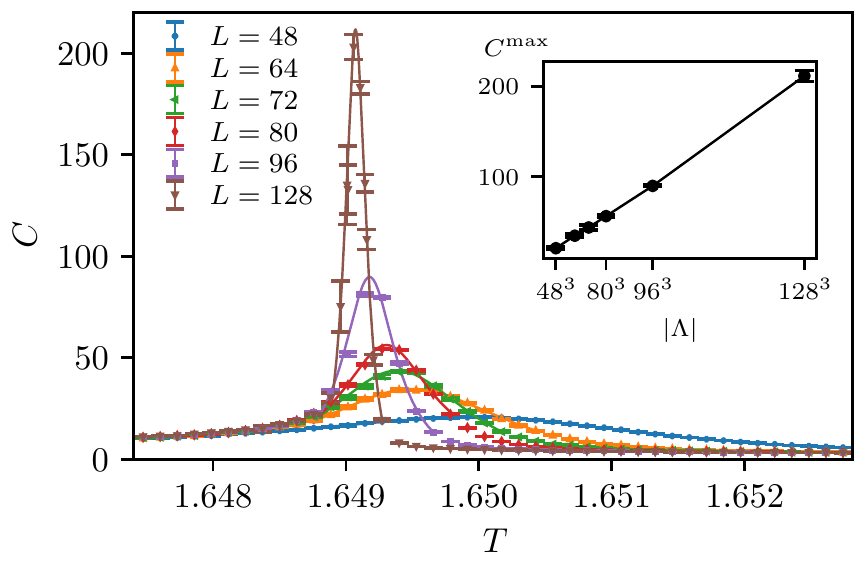}
    \caption{Temperature dependence of the specific heat $C$ for different system sizes at $u=\cot(3\pi/8)$ from QMC. The inset shows the scaling of the maximum  $C^\text{max}$ with system size, extracted from the shown Lorentzian interpolations.}
    \label{fig:C}
\end{figure}

Next, we consider the phase transition. 
We demonstrate that in contrast to earlier claims, the planar spin-nematic order melts across a (weakly) first-order thermal transition. A basic quantity for this purpose is the specific heat $C$, the $T$ dependence of which is shown in Fig.~\ref{fig:C}. For sufficiently large systems, we clearly identify a prominent peak with a scaling $C^\text{max}\propto |\Lambda|$, characteristic of a first-order transition. From an extrapolation of the peak position~\cite{SM}, we obtain the estimate $T_c=1.64900(1)$ for the transition temperature at this parameter value.

Further evidence for the first-order character of the transition is obtained from considering the nematic histogram at $T_c$. This is shown in the inset of Fig.~\ref{fig:PQ}, and exhibits the coexistence of two 
contributions: (i) a broad low-$T$ contribution akin to the one in the main panel, and  (ii) a further, comparably sharp peak near $Q=0$, i.e.,  related to disordered states. The latter emerges only  mildly upon increasing the system size, but  it is clearly resolved for $L\gtrsim 100$. This indicates the rather weak first-order character of the transition.  Histograms based on the internal energy also support this conclusion~\cite{SM}. 

Another quantity that exhibits genuine behavior at first-order transitions is the Binder cumulant $U_Q$, shown in Fig.~\ref{fig:U} across the  transition region. Two properties are noticeable: (i) $U_Q$ develops a substantial dip just above $T_c$, which grows and sharpens with increasing $L$, another characteristic feature of first-order transitions~\cite{Binder1987}.  (ii) The data for $U_Q$ from different system sizes exhibit a crossing at $T_c$. We  can calculate the crossing point value $U^c_Q$ as follows from considering the coexistence of ordered and disordered states: Denoting by $\alpha$ the weight of the ordered states at coexistence, such that $\langle \cdot \rangle_{\beta_c}=\alpha \lim_{\beta\to\beta_c^+}\langle \cdot \rangle_{\bL}+(1-\alpha) \lim_{\beta\to\beta_c^-}\langle \cdot \rangle_{\bL}$, we can express $U^c_Q$ in terms of  the previously calculated moments of $Q$ in the nematic phase, taking into account that they vanish in the paramagnetic phase. This gives
$U_Q^c=1-5/(7\alpha)$. 
We finally need to determine the mixing parameter $\alpha$ at the first-order transition in the quantum system described by $\Ham$. A related issue appears for first-order transitions in {\it classical} models with continuous variables, and this has been addressed only recently~\cite{Xu2019}: Based on  the fact that for the discrete $q$-state Potts model the corresponding parameter is given in terms of the number $q$ of distinct degenerate low-$T$ sectors with respect to the single paramagnetic sector by $\alpha=q/(q+1)$, it was argued that for the continuous case, $\alpha$ is obtained upon replacing $q$ in the above formula by the integral measure of the space of extremal states. In the current case this measure is given by the area $2\pi$ of the projective sphere $P\bbS^2$, i.e., $\alpha=2\pi/(2\pi+1)$. The  value of $U_Q^c=2/7 - 5/(14\pi)=0.1720\dots$ resulting from this heuristics indeed matches remarkably well to the QMC data, cf. the inset in Fig.~\ref{fig:U}. This demonstrates that PD calculations  provide an accurate quantitative description of the planar spin-nematic phase of the spin-one quantum magnet. It would  be valuable to base  the heuristics of Ref.~\cite{Xu2019} on more rigorous considerations for both continuous and quantum variables.

\textit{Conclusions.} We  used a combination of QMC  and PD calculations, based on a loop model formulation, to uncover weakly first-order thermal melting transitions of planar spin-nematic states realized in quantum spin-one systems with SU(2)-symmetric interactions. We demonstrated explicitly how generic properties of both the low-temperature nematic phase and the phase coexistence line can be calculated based on the PD conjecture, with remarkable agreement to QMC results.   Further studies, e.g., based on RG approaches,  will be useful in order to explain the weakness of these transitions via the tuning mechanism, or by connecting it to the ideas of walking and fixed-point annihilation within this well-defined framework of a comparably simple quantum spin model. 

\textit{Acknowledgements.}
We thank A. Honecker, J. B. Hauck and P. Sengupta for insightful discussions. We acknowledge support by the Deutsche Forschungsgemeinschaft (DFG) through Grant No. WE/3649/4-2 of the FOR 1807 and  RTG 1995, and thank the IT Center at RWTH Aachen University and  JSC Jülich for access to computing time through  JARA CSD.

\bibliography{paper.bib}

 










\pagebreak
\begin{widetext}
\begin{center}
\textbf{\large Supplemental Materials for "Poisson-Dirichlet distributions and weakly first-order\\ spin-nematic phase transitions"}
\end{center}
\end{widetext}

\setcounter{equation}{0}
\setcounter{figure}{0}
\setcounter{table}{0}
\setcounter{page}{1}
\makeatletter
\renewcommand{\theequation}{S\arabic{equation}}
\renewcommand{\thefigure}{S\arabic{figure}}

\section{Poisson-Dirichlet distributions and split-merge processes}

We introduce  here the family of Poisson-Dirichlet distributions and explain how they relate to   split-merge processes in general. This also helps to understand why these distributions appear in the loop models, and more importantly, it allows us to later calculate the PD parameter $\theta$ that identifies the distribution of loop lengths.

The relevant objects here are partitions $(X_1,X_2\dots)$ of $[0,1]$. That is, the numbers $X_i$ satisfy $X_1 \geq X_2 \geq \dots \geq 0$ and $\sum_{i=1}^\infty X_i = 1$. The simplest definition of Poisson-Dirichlet involves the related Griffiths-Engen-McCloskey (GEM) distribution. The latter is a residual allocation measure built from Beta$(1,\theta)$ random variables. Recall that a Beta$(1,\theta)$ random variable has probability density function $\theta (1-s)^{\theta-1}$, $s \in [0,1]$, where the parameter $\theta$ is positive. Then, with $Y_1, Y_2, \dots$ being independent Beta$(1,\theta)$ random variables, we consider the vector
\[
\bigl( Y_1, (1-Y_1) Y_2, (1-Y_1) (1-Y_2) Y_3, \dots \bigr).
\]
One can check that these positive numbers add up to 1. Rearranging them in decreasing order, one gets a random partition of $[0,1]$ selected with the Poisson-Dirichlet distribution PD$(\theta)$.

We denote $\mathbb E_{{\rm PD}(\theta)}$ the expectation with respect to Poisson-Dirichlet PD$(\theta)$. We will apply it to functions of the form $\prod_{i=1}^\infty f(X_i)$ where $f$ is a bounded function $[0,1] \to \mathbb C$ such that $f(s) = 1 + o(s)$ around $s=0$ --- this guarantees that the infinite product converges, and also that small loops do not contribute. Since the order of partition elements is not important we can directly use the GEM measure. Concretely, this gives
\begin{equation}
\begin{split}
    \mathbb E_{{\rm PD}(\theta)} &\biggl[ \prod_{i=1}^\infty f(X_i) \biggr] = \biggl( \prod_{i=1}^\infty \int_0^1 \theta (1-s_i)^{\theta-1} d s_i \biggr) \\
    &f(s_1) f \bigl( (1-s_1) s_2 \bigr) f \bigl( (1-s_1) (1-s_2) s_3 \bigr) \dots
\end{split}
\end{equation}
If the function $f$ has the Taylor series $f(s) = 1 + \sum_{k\geq1} a_k s^k$, the expectation above can be computed with the help of the moments formula obtained in \cite{Nahum2013}; we get
\begin{equation}
\label{formula PD}
\begin{split}
&\bbE_{{\rm PD}(\theta)} \biggl[ \prod_{i=1}^\infty f(X_i) \biggr] \\
&= \sum_{n=0}^\infty \frac1{n!}  \sum_{k_1,\dots,k_n=1}^\infty a_{k_1} \dots a_{k_n} \frac{\theta^{n} \, \Gamma(\theta) \, \Gamma(k_{1}) \dots \Gamma(k_{n})}{\Gamma(\theta + k_{1} + \dots + k_{n})}.
\end{split}
\end{equation}
See \cite[Eq.\ (4.16)]{Ueltschi2017}.

The split-merge process (also called coagulation-fragmentation) is a Markov process on partitions of $[0,1]$. Each step consists of either merging two distinct elements, or splitting in two a given element (in which case it is split uniformly). Let $g_m,g_s$ be two positive parameters.
In its continuous-time version, the partition elements $X_i, X_j$ ($i \neq j$) are merged at rate $2 g_m X_i X_j$; the element $X_i$ is split at rate $g_s X_i^2$. The invariant measure is Poisson-Dirichlet with parameter $\theta = g_s/g_m$ \cite{Goldschmidt2011,Tsilevich2000,Diaconis2004}.

\section{Identifying the Poisson-Dirichlet parameter}

We first present a Markov process that has the measure obtained from the loop model representation of the quantum partition function $Z$ as the invariant measure. Based on this formulation, we can then obtain the PD parameter $\theta$.

It is convenient  to first discretise the "time" interval $[0,\beta]$ with mesh $1/n$. Given a realisation $\omega$ of crosses and double bars, let $C(\omega)$ and $B(\omega)$ denote the number of crosses and double bars, respectively. On an arbitrary finite lattice $\Lambda$ with set of bonds $\caB_\Lambda$, the measure can be written as
\begin{equation}
\label{measure on omega}
\begin{split}
  \mu(\omega) = & \frac1Z 3^{|\caL(\omega)|} \bigl( \tfrac{u}n \bigr)^{C(\omega)} \bigl( \tfrac{1-u}n \bigr)^{B(\omega)} \\
  & \times \bigl( 1 - \tfrac1n \bigr)^{|\caB_\Lambda| \beta n - C(\omega) - B(\omega)}.
\end{split}
\end{equation}

Let $R(\omega,\omega')$ denote the transition matrix $\omega \mapsto \omega'$, the detailed balance equation is
\begin{equation}
\begin{split}
    &3^{|\caL(\omega)|} \left( \tfrac n{u} \right)^{C(\omega)} \left( \tfrac{1-u}n \right)^{B(\omega)} R(\omega,\omega') = \\
    &3^{|\caL(\omega')|} \left( \tfrac n{u} \right)^{C(\omega')} \left( \tfrac{1-u}n \right)^{B(\omega')} R(\omega',\omega).
\end{split}
\end{equation}
Here is a natural process that satisfies the equation above:
\begin{itemize}
\item A new cross appears in $\{i,j\} \times [t,t+\frac1n]$ at rate $\sqrt3 \frac{u}n$ if it causes a loop to split; at rate $\frac1{\sqrt3} \frac{u}n$ if it causes two loops to merge; at rate $\frac{u}n$ if the number of loops does not change.
\item Same with double bars, but with $1-u$ instead of $u$.
\item An existing cross or double bar is removed at rate $\sqrt3$ if its removal causes a loop to split; at rate $\frac1{\sqrt3}$ if its removal causes two loops to merge; at rate 1 if the number of loop remains constant.
\end{itemize}
Notice that any new cross or double bar between two loops causes them to merge. When $u \in (0,1)$, a subtle phenomenon occurs: a new cross or double bar may either cause it to split, or reorganise it without splitting it (this is akin to $0 \leftrightarrow 8$); either occurs with probability $\frac12$.

Let $\gamma, \gamma'$ be two macroscopic loops of lengths $\ell(\gamma), \ell(\gamma')$. They are spread all over $\Lambda$ and they interact between one another, and among themselves, in an essentially mean-field fashion. There exists a constant $c_{1}$ such that a new cross or double bar that causes $\gamma$ to split, appears at rate $\tfrac14 \sqrt3 \, c_{1} \frac{\ell(\gamma)^{2}}{\beta |\Lambda|}$; a new cross or double bar that causes $\gamma$ and $\gamma'$ to merge appears at rate $(c_{1} / \sqrt3) \frac{\ell(\gamma) \ell(\gamma')}{\beta |\Lambda|}$. There exists another constant $c_{2}$ such that the rate for an existing cross or double bar to disappear is $\tfrac14 \sqrt\theta \, c_{2}  \frac{\ell(\gamma)^{2}}{\beta |\Lambda|}$ if $\gamma$ is split, and $(c_{2} / \sqrt3) \frac{\ell(\gamma) \ell(\gamma')}{\beta |\Lambda|}$ if $\gamma$ and $\gamma'$ are merged. Consequently, $\gamma$ splits at rate
\begin{equation}
    \tfrac14 \sqrt3 (c_{1}+c_{2}) \frac{\ell(\gamma)^{2}}{\beta |\Lambda|} \equiv \tfrac12 r_{\rm s} \ell(\gamma)^{2}
\end{equation}
and $\gamma, \gamma'$ merge at rate
\begin{equation}
    \frac1{\sqrt3} (c_{1}+c_{2}) \frac{\ell(\gamma) \ell(\gamma')}{\beta |\Lambda|} \equiv r_{\rm m} \ell(\gamma) \ell(\gamma').
\end{equation}
Because of effective averaging over the whole domain, the constants $c_{1}$ and $c_{2}$ are the same for all loops and for both the split and merge events. This key property is certainly not obvious and the interested reader is referred to a detailed discussion for lattice permutations with numerical checks \cite{Grosskinsky2012}. It follows that the lengths of macroscopic loops satisfy an effective split-merge process, and the invariant distribution is Poisson-Dirichlet with parameter $\theta = r_{\rm s} / r_{\rm m} = 3/2$ \cite{Tsilevich2000,Diaconis2004,Goldschmidt2011}.

For $u = 0$ or $u = 1$, the "subtle phenomenon" above does not occur; splits then happen at twice the rate, and the Poisson-Dirichlet parameter is $\theta = 3$.

\section{Poisson-Dirichlet calculation of "nematic histogram"}

We study the distribution of the operator $Q=\frac1{|\Lambda|} \sum_{i \in \Lambda} Q_i$ in the Gibbs state $\langle \cdot \rangle_{\bL}$. To be precise, we seek to identify the density $\rho_Q$ such that for any function $g$, we have
\begin{equation}
    \langle g( Q ) \rangle_{\bL} = \int_{-\infty}^\infty \rho_Q(s) g(s) ds.
\end{equation}
Choosing $g(s) = e^{i ks}$ gives the characteristic function of $\rho_Q$. Happily, we can use Eq.\ (3) from the main text to get an expression that involves the lengths of the loops. The Poisson-Dirichlet conjecture states that, as $\Lambda \to \bbZ^3$, we can replace the expectation in the loop model by the expectation with respect to PD(3/2), scaled by a number $\eta = \eta(\beta)$ that represents the fraction of sites in long loops at imaginary time 0 ($\eta \in [0,1]$). We then get
\begin{equation}
\begin{split}
    \lim_{L\to\infty} \langle e^{ikQ} \rangle_{\bL} 
    &= \bbE_{{\rm PD}(\frac32)} \biggl[ \prod_{j=1}^\infty \bigl( \tfrac13 e^{-\frac23 ik \eta X_j} + \tfrac23 e^{\frac13 ik \eta X_j} \bigr) \biggr] \\
    &= e^{-\frac23 ik \eta} \, \bbE_{{\rm PD}(\frac32)} \Bigl[ \prod_{j\geq1} \bigl( \tfrac13 + \tfrac23 e^{ik \eta Y_j} \bigr) \Bigr].
\end{split}
\end{equation}
We can use Eq.\ \eqref{formula PD} and we get (see \cite{Ueltschi2017} for more details)
\begin{equation}
\label{result PD}
    \lim_{L\to\infty} \langle e^{ikQ} \rangle_{\bL}= e^{-\frac23 ik \eta} \, \Gamma(\tfrac32) \sum_{r=0}^\infty \frac{(ik \eta)^r}{\Gamma(r+ \frac32)}.
\end{equation}
We calculate below its inverse Fourier transform, see Eq.\ \eqref{eq rho_Q}.

\section{Symmetry breaking calculation of "nematic histogram"}

Let $n^*$ denote the "spontaneous nematisation" in the z direction
\begin{equation}
    n^* = \langle Q_i \rangle_{\vec e_{\rm z}},
\end{equation}
where $i$ is any site. Let $Q_i^{\vec a}$ be the spin rotation of the operator $Q_i$, namely
\begin{equation}
    Q_i^{\vec a} = \bigl( a_1 S_i^{\rm x} + a_2 S_i^{\rm y} + a_3 S_i^{\rm z} \bigr)^2 - \tfrac23.
\end{equation}
Its expectation can be expressed in terms of $n^*$:
\begin{equation}
    \langle Q_i^{\vec a} \rangle_{\vec e_{\rm z}} = \sum_{j={\rm x,y,z}} a_j^2 \langle (S_i^j)^2 - \tfrac23 \rangle_{\vec e_{\rm z}} + \sumtwo{j,k={\rm x,y,z}}{j\neq k} a_j a_k \langle S_i^j S_i^k \rangle_{\vec e_{\rm z}}.
\end{equation}
It is clear that $\langle \cdot \rangle_{\vec e_{\rm z}}$ is invariant under spin rotations around $\vec e_{\rm z}$, and also that $\langle S_i^{\rm z} \rangle_{\vec e_{\rm z}} = 0$, so that $\langle S_i^j S_i^k \rangle_{\vec e_{\rm z}} = 0$ for all $j \neq k$. Further, since $(S_i^{\rm x})^2 + (S_i^{\rm y})^2 + (S_i^{\rm z})^2 = 2$, we have
\begin{equation}
    \langle (S_i^{\rm x})^2 - \tfrac23 \rangle_{\vec e_{\rm z}} =  \langle (S_i^{\rm y})^2 - \tfrac23 \rangle_{\vec e_{\rm z}} = -\tfrac12  \langle (S_i^{\rm z})^2 - \tfrac23 \rangle_{\vec e_{\rm z}}.
\end{equation}
This gives
\begin{equation}
\label{rotated state}
    \langle Q_i^{\vec a} \rangle_{\vec e_{\rm z}} = n^* (a_3^2 - \tfrac12 a_1^2 - \tfrac12 a_2^2).
\end{equation}
This allows to calculate
\begin{equation}
\label{symm break}
\begin{split}
    \lim_{L\to\infty} \langle & e^{ikQ} \rangle_{\bL} = \lim_{L \to \infty} \int_{P\bbS^2} \Bigl\langle e^{\frac{ik}{|\Lambda|} \sum_{i\in\Lambda} Q_i} \Bigr\rangle_{\vec a} d\vec a \\
&= \lim_{L \to \infty} \int_{P\bbS^2} \Bigl\langle e^{\frac h{|\Lambda|} \sum_{i\in\Lambda} Q_i^{\vec a}} \Bigr\rangle_{\vec e_{\rm z}} d\vec a \\
&= \int_{P\bbS^2} e^{ik n^* (a_3^2 - \frac12 a_1^2 - \frac12 a_2^2)} d\vec a \\
&= e^{ik n^*} \int_0^{\pi/2} d\theta \sin\theta \, e^{-\frac32 ik n^* \sin^2 \theta}.
\end{split}
\end{equation}
Expanding the exponential in Taylor series and calculating the trigonometric integrals, we obtain
\begin{equation}
\label{result symm break}
    \lim_{L\to\infty} \langle e^{ikQ} \rangle_{\bL} = e^{ik n^*} \Gamma(\tfrac32) \sum_{r=0}^\infty \frac{(-\frac32 ik n^*)^r}{\Gamma(r+\frac32)}.
\end{equation}
We recover the result in Eq.\ \eqref{result PD} provided that
\begin{equation}
\label{value n*}
    n^* = -\tfrac23 \eta.
\end{equation}

It is worth pointing out that $n^*$ is negative. This allows to understand the nature of the nematic extremal states. Indeed, a natural candidate is the "axial nematic" state
\begin{equation}
    \langle \cdot \rangle_{\vec a} = \lim_{h \to 0+} \lim_{L\to\infty} \langle \cdot \rangle_{\Ham - h \sum_{i \in \Lambda} (\vec a \cdot \vec S_i)^2}.
\end{equation}
One can write a loop representation for the state $\langle \cdot \rangle_{\vec e_{\rm z}}$ where short loops have spin values $-1,0,+1$ and long loops have spin values $-1,+1$. The nematic order parameter would then be equal to
\begin{equation}
    \tilde n = \lim_{L\to\infty} \langle Q \rangle_{\vec e_{\rm z}} = \tfrac13 \eta.
\end{equation}
This contradicts Eq.\ \eqref{value n*}. Instead, it turns out that extremal states are "planar nematic":
\begin{equation}
\label{extremal state}
    \langle \cdot \rangle_{\vec a} = \lim_{h \to 0+} \lim_{L\to\infty} \langle \cdot \rangle_{\Ham + h \sum_{i \in \Lambda} (\vec a \cdot \vec S_i)^2}.
\end{equation}
(Notice the "$+$" sign in front of $h$). In its loop representation, long loops have the spin value 0, and Eq.\ \eqref{value n*} holds true. The fact that extremal states are planar nematic was pointed out in \cite{Fridman2013}.

We can calculate the density $\rho_Q$ starting from Eq.\ \eqref{symm break}.
\begin{equation}
\begin{split}
    \rho_Q(s) &= \frac1{2\pi} \int_{-\infty}^\infty dk \, e^{-isk} e^{-\frac23 ik \eta} \int_0^{\pi/2} d\theta \, \sin\theta \, e^{ik \eta \sin^2\theta} \\
    &= \frac1{2\pi} \int_0^{\pi/2} d\theta \, \sin\theta \int_{-\infty}^\infty dk \, e^{ik (-s -\frac23 \eta + \eta \sin^2\theta)} \\
    &= \int_0^{\pi/2} d\theta \, \sin\theta \, \delta(\eta \sin^2\theta -s -\tfrac23 \eta) \\
    &= \frac1{2\eta} \int_0^\eta \frac{dt}{\sqrt{1 - \frac t\eta}} \delta(t - s -\tfrac23 \eta).
\end{split}
\end{equation}
We used the change of variables $t = \eta \sin^2 \theta$. We finally obtain the density for the nematic observable:
\begin{equation}
\label{eq rho_Q}
     \rho_Q(s) = \begin{cases} \frac1{2\sqrt\eta \sqrt{\frac13 \eta - s}} & \text{if } -\frac23 \eta \leq s \leq \frac13 \eta, \\ 0 & \text{otherwise.} \end{cases}
\end{equation}

\section{Binder cumulants}

We can use the loop representation to get expressions for the moments of the operator $Q = \frac1{|\Lambda|} \sum_{i\in\Lambda} Q_i$; then we use the Poisson-Dirichlet conjecture to write them all in terms of a single unknown variable, the fraction of sites in long loops $\eta$. The Binder cumulants follow, and they are given by ratios that do not depend on $\eta$ any more.
We write the calculations in some details since they cannot be found in the literature, to the best of our knowledge.

Here are identities that are exact in the infinite-volume limit:
\begin{equation}
    \begin{split}
        &\langle Q^2 \rangle_{\bL} = \tfrac29 \bbP_{\bL}^{\rm loops} [ i_1, i_2 \text{ in same loop}], \\
        &\langle Q^3 \rangle_{\bL} = -\tfrac2{27} \bbP_{\bL}^{\rm loops} [ i_1, i_2, i_3 \text{ in same loop}].
    \end{split}
\end{equation}
Here $i_1, i_2, i_3$ (and $i_4$ below) are sites that are very distant from one another. The first identity can be found in \cite{Ueltschi2013}; the second identity is similar. As for the 4th moment, we have
\begin{equation}
    \begin{split}
    &\langle Q^4 \rangle_{\bL} = \tfrac2{27} \bbP_{\bL}^{\rm loops} [ i_1, i_2, i_3, i_4 \text{ in same loop}] \\
    &+ \tfrac4{27} \bbP_{\bL}^{\rm loops} [ i_1, i_2 \text{ in same loop}, i_3, i_4 \text{ in other loop}].
    \end{split}
\end{equation}
Since the sites are distant, it is necessary that they belong to long loops in order to have a chance to be in the same loop. We can then use the Poisson-Dirichlet conjecture and we get
\begin{equation}
    \begin{split}
    &\bbP_{\bL}^{\rm loops} [ i_1, \dots, i_n \text{ in same loop}] \\
    &= \eta^n \bbP_{{\rm PD}(\theta)} [n \text{ random points in same partition element}].
    \end{split}
\end{equation}
The latter is the probability that, if we choose a random partition of $[0,1]$ according to PD($\theta$), and $n$ independent points in $[0,1]$, all $n$ points find themselves in the same partition element. This does not depend on the order of the elements so we can replace the Poisson-Dirichlet distribution by the GEM distribution. We calculate it by summing over the probability that the $n$ random points belong to the $k$th element; namely,
\begin{equation}
    \begin{split}
        &\bbP_{\bL}^{\rm loops} [ i_1, \dots, i_n \text{ in same loop}] \\
        &= \eta^n \sum_{k=1}^\infty \bbP_{{\rm PD}(\theta)} [n \text{ points in $k$th partition element}] \\
        &= \eta^n \sum_{k=1}^\infty \bbE_{\{Y_i\}}[(1-Y_1)^n \dots (1-Y_{k-1})^n Y_k^n].
    \end{split}
 \end{equation}
Since the $\{Y_i\}$ are independent, the expectation factorises and we get       
\begin{equation}
    \begin{split}
        \bbP_{\bL}^{\rm loops} [ i_1, \dots, i_n & \text{ in same loop}] = \eta^n \frac{\bbE_{{\rm Beta}(1,\theta)}[Y^k]}{1 - \bbE_{{\rm Beta}(1,\theta)}[(1-Y)^k]} \\
        &= \eta^n \frac{\Gamma(1+\theta) \Gamma(n)}{\Gamma(n+\theta)}.
    \end{split}
\end{equation}
A similar calculation gives
\begin{equation}
\begin{split}
&\bbP_{\bL}^{\rm loops} [i_1, i_2 \text{ in same loop}, i_3, i_4 \text{ in other loop}] \\
&= 2\eta^4 \sum_{k<\ell} \bbP_{{\rm PD}(\theta)} [i_1, i_2 \text{ in $k$th element}, i_3, i_4 \text{ in $\ell$th el.}] \\
&= 2\eta^4 \sum_{k,\ell=1}^\infty \bbE_{\{Y_i\}}[(1-Y_1)^4 \dots (1-Y_{k-1})^4 \\
& \hspace{25mm} \cdot Y_k^2 (1-Y_k)^2 \dots (1-Y_{k+\ell-1})^2 Y_{k+\ell}^2] \\
&= 2\eta^4 \frac{\bbE_{{\rm Beta}(1,\theta)}[Y^2 (1-Y)^2] \bbE_{{\rm Beta}(1,\theta)}[Y^2]}{(1-\bbE_{{\rm Beta}(1,\theta)}[(1-Y)^4]) (1-\bbE_{{\rm Beta}(1,\theta)}[(1-Y)^2])} \\
&= 2\eta^4 \frac{\theta \Gamma(1+\theta)}{\Gamma(4+\theta)}.
\end{split}
\end{equation}

Combining the terms above we get the moments for the nematic phase ($\theta = 3/2$) and for the SU(3) phases ($\theta=3$) given in Tab. I of the main text. Alternatively we could have looked at the Taylor series of $\langle e^{i k Q} \rangle_\beta$ from the expression in Eq.~\eqref{expression for exp(ikQ)}.

We now calculate the moments using symmetry breaking; we express them in terms of $n^*$. The $k$th moment is given by
\begin{equation}
    \begin{split}
        \langle Q^k \rangle_{\bL} &= \frac1{|\Lambda|^k} \sum_{i_1,\dots,i_k} \langle Q_{i_1} \dots Q_{i_k} \rangle_{\bL} \\
        &=\frac1{|\Lambda|^k} \sum_{i_1,\dots,i_k} \int_{P\bbS^2} d\vec a \, \langle Q_{i_1} \rangle_{\vec a} \dots \langle Q_{i_k} \rangle_{\vec a}.
    \end{split}
\end{equation}
We used the fact that extremal states are "clustering" and that for $\Lambda$ large, the main contribution in the sum comes from distant sites. We now use translation invariance and we rotate the observable rather than the state, so as to get
\begin{equation}
    \begin{split}
        \langle Q^k \rangle_{\bL} &= \int_{P\bbS^2} d\vec a \, \langle Q_i^{\vec a} \rangle_{\vec e_{\rm z}}^k \\ &= (n^*)^k \int_{P\bbS^2} d\vec a \, (a_3^2 - \tfrac12 a_1^2 - \tfrac12 a_2^2)^k \\
        &= (n^*)^k \int_0^{\pi/2} d\theta \, \sin\theta \, (\tfrac32 \cos^2\theta - \tfrac12)^k.
    \end{split}
\end{equation}
We used Eq.\ \eqref{rotated state} to get the second line. Calculating the integral we finally get
\begin{equation}
    \begin{split}
        &\langle Q^2 \rangle = \tfrac15 (n^*)^2, \\
        &\langle Q^3 \rangle = \tfrac2{35} (n^*)^3, \\
        &\langle Q^4 \rangle = \tfrac3{35} (n^*)^4.
    \end{split}
\end{equation}
This is compatible with the values in the Table \ref{Table:Qn} if we assume validity of the relation \eqref{value n*}. Notice that $n^*$ happens to be negative.

The calculation using symmetry breaking is simpler than that with Poisson-Dirichlet. However, the heuristics is more subtle and the result may be uncertain. For $u=0$ and $u=1$, the Poisson-Dirichlet calculations can be carried out without much hesitation (with $\theta=3$) but symmetry breaking is not immediate.

\medskip

\section{Energy histograms}
\begin{figure}[t]
    \centering
    \includegraphics{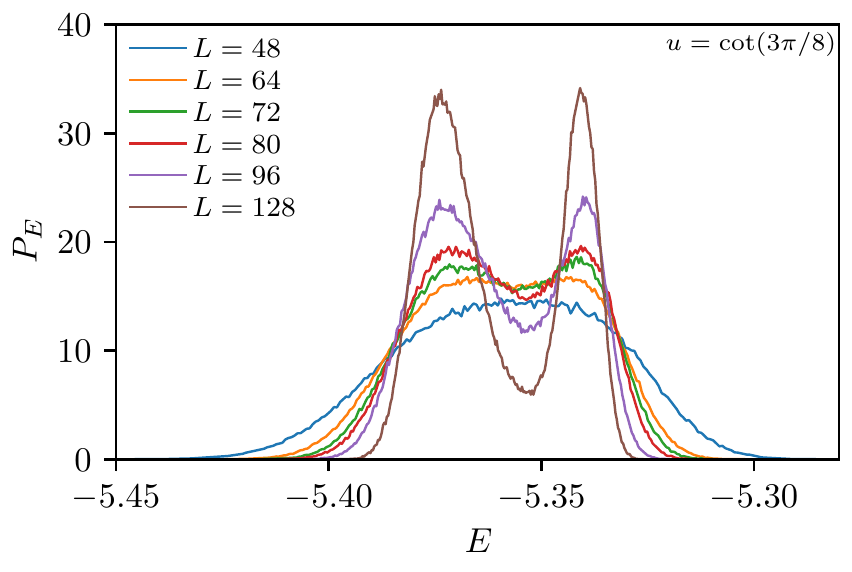}
    \caption{Energy histograms $P_E$ for $u=\cot(\phi=3\pi/8)=0.41412...$ for various system sizes $L$ at temperatures with equal peak height.}
    \label{figShist}
\end{figure}

\begin{figure}[t]
    \centering
    \includegraphics{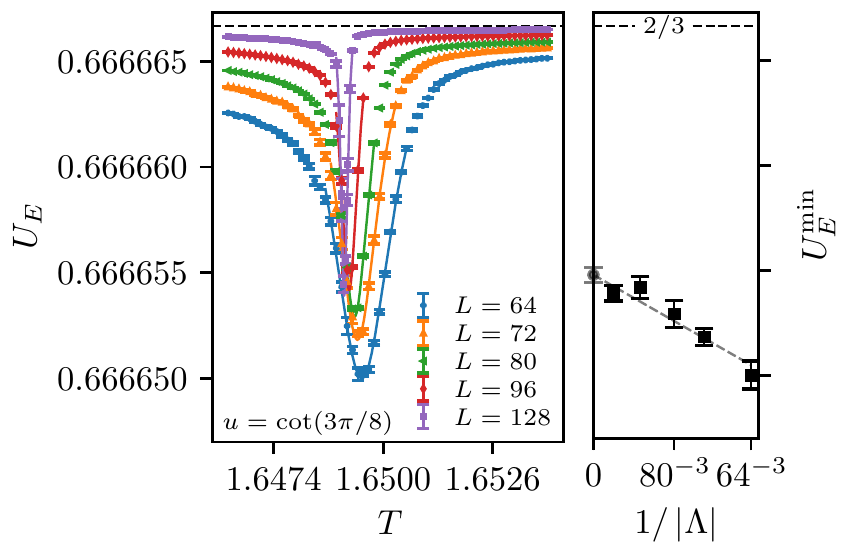}
    \caption{Left panel: Cumulant $U_E$ of the energy distribution as a function of temperature near $T_c$ at $u=\cot(\phi=3\pi/8)=0.41412...$ for various system sizes $L$. Right panel: Finite-size extrapolation of the minimum value $U_E^\text{min}$ of $U_E$. }
  \label{figSUE}
\end{figure}
Further evidence for the first-order character of the planar spin nematic melting transition is obtained from analysing  energy histograms $P_E$ near the transition temperature. Within the stochastic series expansion QMC approach, the energy histogram $P_E$ is readily available from the histogram of the expansion order~\cite{Sandvik1999}. We  obtain histograms with a pronounced two-peak structure for sufficiently large system sizes, indicative of phase coexistence. In particular, 
we can use standard histogram-reweighting~\cite{Ferrenberg1988} in order to access the energy histograms at any temperatures $T$ in the vicinity of a base temperature, at which the QMC simulations were actually performed. For each system size, this base temperature was taken from the peak position of the specific heat. This reweighting approach allows us to adjust $T$ such as to obtain histograms $P_E$ with an equal peak height of the two peaks~\cite{Lee1990}. These are shown in Fig.~\ref{figShist} for our reference value of $u=\cot(\phi=3\pi/8)=0.41412...$. We identify a pronounced two-peak structure for $L\gtrsim 64$. While the dip for $L=64$ is still shallow, it becomes  deeper for increasing values of $L$, in agreement with the predictions by Binder~\cite{Binder1987} and Lee and Kosterlitz~\cite{Lee1990,Lee1991} for a first-order transition. The fact that the minimum takes on a substantial value even for $L=128$ reflects the fact that the transition is  weakly first-order.

\begin{figure}[t]
    \centering
    \includegraphics{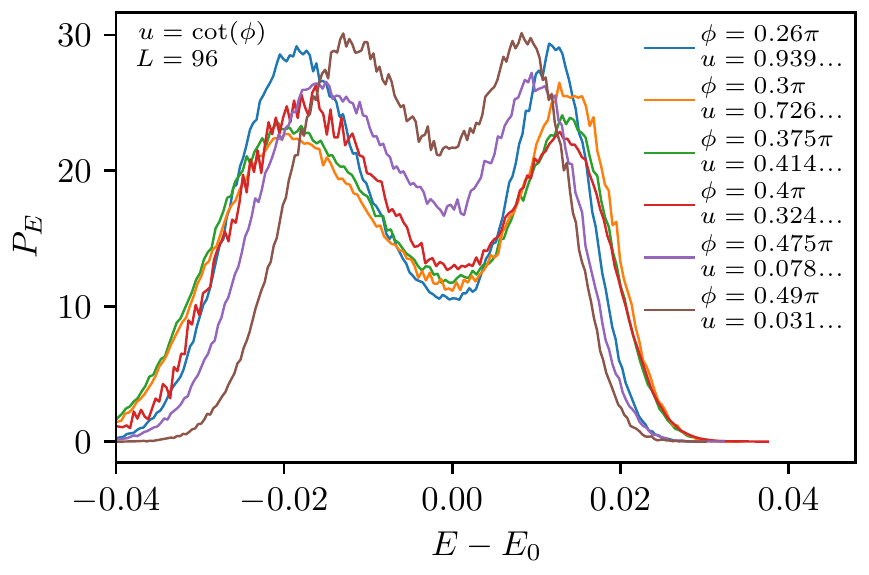}
    \caption{Energy histograms $P_E$ for various values of $u$ for $L=96$ at temperatures with equal peak height. For better comparison the individual histograms are shown shifted with respect to the energy $E_0$ of the minimum.}
    \label{figShistvar}
\end{figure}

\begin{figure}[t]
    \centering
    \includegraphics{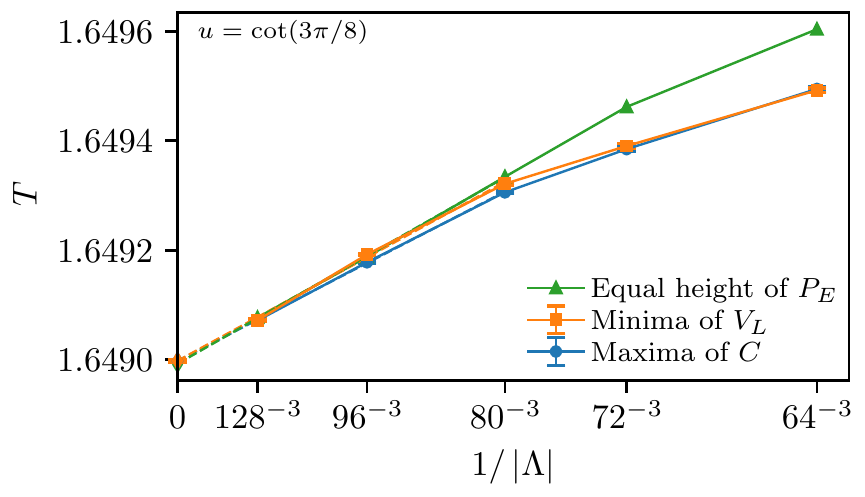}
    \caption{Finite-size extrapolation of different estimators for the transition temperature $T_c$ at $u=\cot(\phi=3\pi/8)=0.41412...$.}
    \label{figSTc}
\end{figure}

We also analyzed the fourth-order cumulant
\begin{equation}
    U_E=1-\frac{1}{3}\frac{\langle H^4\rangle_\beta}{\langle H^2\rangle_\beta^2}
\end{equation}
of the energy distribution in the vicinity of $T_c$. This is shown for different system sizes in Fig.~\ref{figSUE}. We observe a narrow dip in $U_E$ at a temperature that approaches $T_c$ upon increasing $L$. While the minimum value  $U_E^\text{min}$ tends towards $2/3$ for increasing $L$, a finite-size extrapolation with a $1/L^3$-scaling (cf. the right panel of  Fig.~\ref{figSUE}) shows that in the thermodynamic limit $U_E$  remains well below $2/3$. This is another strong indication for the first-order character of the phase transition~\cite{Janke1993}.

Thus far, we concentrated on $u=\cot(\phi=3\pi/8)=0.414...$, but we also performed QMC simulations at different values of $u$ across the planar spin nematic regime. Energy histograms  $P_E$ for several values of $u=\cot(\phi)$ are shown in Fig.~\ref{figShistvar}. Here, the energy has been shifted with respect for the dip position, denoted $E_0$, for better comparison. The energy histograms $P_E$  all exhibit a characteristic two-peak structure. We furthermore find that upon approaching the $u=0$ ($\phi=\pi/2$) end point of the planar spin nematic phase, the relative value of the minimum between the peaks increases slightly. For example, at $u=0.414...$ the ratio between the  the local minimum value of $P_E$ and its maximum value is about 0.51, while at $u=0.031..$, this ratio has increased  to about 0.71.
This indicates that the transition becomes even weaker first-order upon approaching this SU(3) point. Moving towards the other end point of the planar spin nematic phase at $u=1$ ($\phi=\pi/4$), we do not observe a similar weakening of the first-order character of the nematic transition. Previous work on the thermal transitions out of the ferromagnetic phase at both SU(3) points claims that both transitions are continuous~\cite{Harada2002}. We note that on the $L=96$ system size, we similarly were not able to resolve any two-peak structure in $P_E$ at the SU(3) point  at $u=1$. 
It would thus certainly be interesting to further examine the SU(3) points in more detail in future work  (where  the low-$T$ phase is ferromagnetically ordered) on even larger lattices than accessible to us, in order to assess the  conclusion of Ref.~\cite{Harada2002} regarding the nature of the  phase transitions at the  SU(3) points. 

\section{Determination of $T_c$}
Here, we detail the estimation of the transition temperature $T_c$, focusing again on our reference value  $u=\cot(\phi=3\pi/8)$=0.41412..... We consider three different estimators for $T_c$, obtained upon performing an extrapolation to the thermodynamic limit of (i) the position of the maximum in the specific heat $C$, (ii)  the position of the minimum in $U_E$, and (iii) the temperature for which the peaks in the energy histogram $P_E$ have equal height. As shown in Fig.~\ref{figSTc}, all three quantities extrapolate with a $1/L^3$-scaling for large system sizes to a mean estimate $T_c=1.64900(1)$ for the transition temperature, as quoted  in the main text.  

\end{document}